# New front and back-end electronics for the upgraded GABRIELA detection system

K. Hauschild, R. Chakma, A. Lopez-Martens, K. Rezynkina, V. Alaphillipe, L. Gibelin, N. Karkour, and D. Linget

*CSNSM, IN2P3-CNRS, UMR 8609, F-91405, Orsay, France*
†*E-mail: karl.hauschild@csnsm.in2p3.fr*
*https://www.csnsm.in2p3.fr/GABRIELA*

A.V. Yeremin, A.G. Popeko, O.N. Malyshev, V.I. Chepigin, A.I. Svirikhin, A.V. Isaev, E.A. Sokol, M.L. Chelnokov, Yu.A. Popov, D.E. Katrasev, A.N. Kuznetsov, A.A. Kuznetsova, M.S. Tezekbayeva

*FLNR, JINR, 141980 Dubna, Russia*

O. Dorvaux, B.J.P. Gall, P. Brionnet, K. Kessaci, C. Mathieu

*IPHC-DRS/ULP, IN2P3-CNRS, F-67037, Strasbourg,*

The GABRIELA [1] set-up is used at the FLNR to perform detailed nuclear structure studies of transfermium nuclei. Following the modernization of the VASSILISSA separator (SHELS) [2] the GABRIELA detection system has also been upgraded. The characteristics of the upgraded detection system will be presented along with results from some recent electronics tests.

*Keywords*: Template; Proceedings; World Scientific Publishing.

## 1. Introduction

At the Flerov Laboratory for Nuclear Reactions (FLNR) experiments to perform detailed nuclear structure studies of transfermium nuclei are carried out using the GABRIELA setup (**G**amma **A**lpha **B**eta **R**ecoil **I**nvestigations with the **EL**ectromagnetic **A**nalyser) [1]. In 2006 the French Agence Nationale de la Recherche (ANR) awarded grant to modernize the VASSILISSA separator which then became SHELS: the **S**eparator for **H**eavy **EL**ement **S**pectroscopy [2]. This crystallized additional funds from the JINR. Since the recoil implantation detector at the focal plane of SHELS is larger than that used for VASSILISSA ($100\times100$ mm$^2$ cf $60\times60$ mm$^2$) the whole detection system of GABRIELA had to be rescaled. A second ANR grant and further funds from the FLNR were used for this purpose. Not only has the detection hardware been upgraded, but, we are also in the process of choosing new front-end electronics for the silicon detectors and implementing a fully digital DAQ using state of the art digitizers from National Instruments (NI) [3]. These developments will improve both the timing resolution (allowing shorter lived isomeric states to be studied) and the energy resolution (improving the signal-to-background ratio allowing weaker channels to be observed). The characteristics of the upgraded detection system will be presented along with results from some of the electronics tests. In particular, the ability to discriminate 'escape' alpha particles from Internal Conversion Electrons (ICE) in the 'tunnel'. The new electronics will also allow us to detect fast pile-up events in the implantation Double-sided Silicon Strip Detector (DSSD) and perform time of flight measurements between the Micro-Channel Plate (MCP) and the DSSD when only one emissive foil is used (which is crucial to a successful experimental program using asymmetric reactions such as $^{22}$Ne+$^{238}$U and $^{18}$O+$^{242}$Pu).

## 2. GABRIELA and SHELS at the FNLR

The GABRIELA upgrade was completed in the first half of 2016. It consists of a $100\times100\times0.5$ mm$^3$ DSSD implantation detector at the focal plane of SHELS. This DSSD is surrounded by eight $50\times60\times0.7$ mm$^3$ silicon detectors in the backward hemisphere forming a square sectioned tunnel with two detectors on each side. One hyper-pure germanium (Ge) "clover" detector (CLODETTE) is placed in a collinear geometry just behind the recoil implantation DSSD and four single crystal coaxial Ge detectors are placed opposite each face of the tunnel [2,4]. New BGO Compton-suppression shields were developed for both Ge detector types. In front of the germanium detectors the vacuum chamber has 1 mm thin Dural windows to maximize L X-ray transmission. The single crystal detectors also have C-fiber entrance windows ~0.7 mm thick.

## 3. Electronics Tests

In moving from analogue to digital back-end electronics we expect to achieve the following goals: 1) reduced dead time (useful for measuring short lived isomers), 2) improved event timing (from the current micro second level to sub-nano second), 3) improved energy resolution and 4) to be able to distinguish Internal Conversion Electrons from degraded escape alpha particles detected in the tunnel detectors. An additional requirement was not to need VHDL experts in order to program the FPGA that we hope will facilitate on site maintenance. Currently there are only two options available: one from NUTAQ (formally LYRTEC, used to instrument JUROGAM2) and the other from NI. A summary of the characteristics of the digitizers tested is given in Table 1. While NUTAQ [5] can provide 768 channels in a chassis at a competitive price it was unclear that the FPGA was large enough to deal with our (future) signal processing needs. Included in the table are values for NUMEXO2 [7], which will be used for SIRIUS@S3 [8], for which we know that the size of the FPGA is problematic. The solution provided by NI should have a large enough FPGA for future developments. Additionally, we found the NI graphical interface to program the FPGA more user friendly (when the comparison was made). A number of tests were carried out at the CSNSM, IPHC and the FLNR to ensure that the future digital system would perform at least as well as our current analogue system. We report on some of these in the following sections.

Table 1. Pertinent characteristics of the digitizers considered.

| Digitizer | NUTAQ Perseus 611x + M125 | National Instruments PXIe-5170R | NUMEXO2 |
|---|---|---|---|
| Channels/module | 64 | 8 | 16 |
| Modules/chassis | 10 | 18 | 12 |
| FPGA | Virtex-6 550T[a] | Kintex-7 325T | Virtex-6 130T |
| Logic cells per channel | 8.6 k | 40.6 k | 8.0 k |
| RAM | 4 GB | 1.5 GB | |
| Input voltage range(s) | ±1 V | ±0.1, ±0.2, ±1, ±2.5V | ±1, ±4V |
| Sampling frequency | 125 MHz | 250 MHz | 200 MHz |
| ENOB[b] (at 125 MHz) | 11.3 | 11.5 | - |
| ENOB (at 200 MHz) | - | - | 10.6 |
| ENOB (at 250 MHz) | - | 11.1 | - |

[a]The largest compatible FPGA available.
[b]ENOB: Effective Number Of Bits can be defined using the IEEE standard 1057 as
ENOB = $\log_2\{$ [full−scale input range]/[ADC RMS noise x sqrt(12)] $\}$

### 3.1. *MCP-MCP Time-Of-Flight*

Our time-of-flight (ToF) system typically uses two thin emissive foils (EF). Electrons emitted from the foils, following the passage of a heavy ion, are focused on to four MCP detectors by a magnetic field. The set-up is shown schematically in Figure 1(a). The preamplifier (PA) output was passed through an SBLP-39 low-pass filter from minicircuits to increase the rise-time of the signals to be compatible with the 4 ns sampling period of the NI-PXIe-5170R. A passive RF splitter and a 2 ns delay was used to obtain two identical signals and hence an effective digitizer rate of 500 MHz. The LabVIEW data acquisition was triggered by the detection of an alpha particle in the DSSD. Analysis was performed off-line on the data written to disc. In Figure 1(b) a representative signal derived for MCP4 is shown from which the zero-crossing point is extracted. The ToF is given by difference between (MCP1.OR.MCP3) and (MCP2.OR.MCP4). The difference between the zero-crossing points for MCP4 and MCP2 as a function of alpha particle energy and the projection onto the time axis are presented in Figure 1 (c) and (d) respectively. Averaged over the three alpha lines the timing resolutions obtained were $\sigma=156(8)$ ps and $139(11)$ ps for the signals recorded with 250 and 500 MHz sampling rates respectively. With the implementation of a new lower noise and larger bandwidth preamplifier and a 1 GHz digitizer we hope to improve on this.

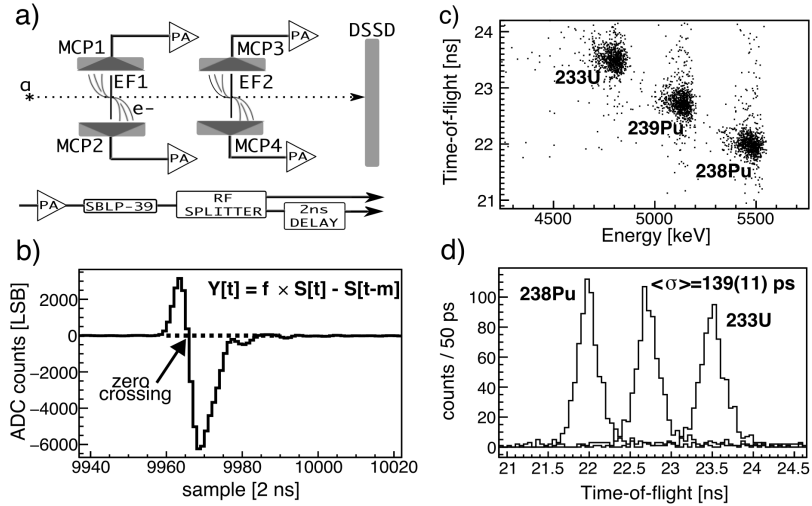

Fig. 1. (a) Schematic of the time-of-flight measurement system (acronyms are explained in the text). (b) A typical digital CFD signal for MCP4. The linearly interpolated zero-crossing point defines the event time. (c) A time-of-flight vs alpha particle energy matrix and (d) the projected time-of-flight.

### 3.2. *MCP-DSSD Time-Of-Flight*

For very asymmetric reactions the foil furthest from the must be removed and the ToF is determined between MCP3 or MCP4 and the DSSD. While having excellent energy resolution for Si detectors (15-18 keV FWHM at 5499 keV and a full-scale of ~250 MeV) [8] our current preamplifiers from TechInvest do not have a fast enough rise-time to achieve sub-ns timing. Two solutions are under investigation: modifications to the preamps developed for the tunnel detectors of SIRIUS and custom preamps from CREMAT. Both can provide impressive energy resolutions as shown in Figure 2(a) for one of the CREMAT preamps used to instrument the implantation detector of GABRIELA. Despite having ~75 cm cables from the vacuum chamber to the preamplifiers we have been able to obtain MCP-DSSD timing resolution, averaged over the three alpha lines, of σ = 800(45) ps in fully digital system. Future tests will be performed with shorter cables (and hence lower parasitic capacitance) in order to improve on these encouraging results.

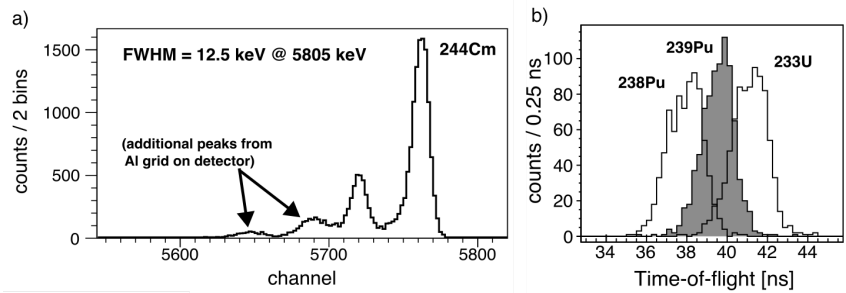

Fig. 2. (a) $^{244}$Cm spectrum measured with the GABRIELA DSSD in situ using a custom built CREMAT preamplifier. (b) MCP4-DSSD time-of-flight for a triple alpha source.

### 3.3. *Particle Discrimination*

Due to the geometry of our setup, sometimes alpha particles from a stronger reaction channel that escape the DSSD and hit the tunnel (leaving energy in both detectors) can mask real fine structure alpha – ICE coincidences that we are trying to study. An example of the ground state alpha decay of $^{254}$No masking $^{255}$No alpha – $^{251}$Fm ICE coincidences is given in Fig. 3(a). Alpha particles and electrons of the same energy have very different ranges in Si. Therefore the charge collection times and hence shape of the preamp signals should be different. Using SHELS we transported $^{233}$U alpha particles from the target position to GABRIELA thus ensuring an e- free source of energy degraded alphas with energies overlapping the ICE from a standard $^{133}$Ba source. In Figure 3(b) it can be clearly seen that ICE and low energy alpha particles can be distinguished from one another purely on the basis of the preamplifier rise-time. To our knowledge this is the first demonstration of ICE/α discrimination at an energy range appropriate to decay spectroscopy. With faster preamplifiers it should be possible to lower the discrimination threshold, but already, it will be possible to remove a large fraction of the escape alpha particles and thus improve the signal-to-noise for the ICE.

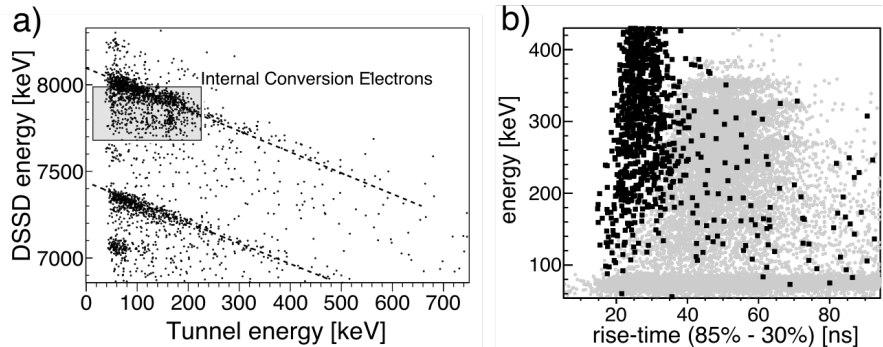

Fig. 3. (a) Prompt coincident DSSD-tunnel energy matrix following the implantation of evaporation residues of the reaction $^{208}$Pb($^{48}$Ca, xn). Dashed lines indicate alpha particles that escape the DSSD and are detected in the tunnel. ICE in $^{251}$Fm from the alpha decay of $^{255}$No are indicated by the shaded box. (b) Particle energy as a function of preamplifier signal rise-time. Black squares: $^{233}$U alphas degraded in energy. Grey circles: conversion electrons from the decay of $^{133}$Ba.

## 4. Conclusions

Using the PXIe-5170R digitizers from National Instruments we have be able to obtain excellent energy resolutions for Ge and Si detectors under experimental conditions: FWHM(Ge) < 1.9 keV at 1332 keV and FWHM < 12.5 keV at 5805 keV. The time-of-flight Measurements, either between two MCPs or MCP-DSSD, can be performed in fully digital system with the required resolution. We have demonstrated that ICE can be discriminated from escape alphas at energies relevant to decay spectroscopy.

## 5. Acknowledgments

The modernization of the VASSILISSA separator, now SHELS, was financed by the French Agence Nationale de la Recerche (Contract No. ANR-06-BLAN-0034-01 and ANR-12-BS05-0013) and by JINR. Work at FLNR was performed partially under the financial support of the Russian Foundation for Basic Research, Contract No. 17-02-00867, 18-52-15004.